\def\bey{\begin{eqnarray}}
\def\eey{\end{eqnarray}}
\def\be{\begin{equation}}
\def\ee{\end{equation}}
\def\ba{\begin{array}}
\def\ea{\end{array}}

\def\gm{\gamma}

\def\ld{\lambda}

\def\af{\alpha}
\def\sg{\sigma}

\def\bt{\beta}

\def\pp{\partial}

\def\ra{\rightarrow}
\def\nnb{\nonumber}
\documentstyle[11pt]{article}

\title{  Energy partition in nonabelian gauge fields with thermodynamic interactions}

\author{W.Z.Jiang\footnote{  Electronic address: jiangwz02@hotmail.com}
\footnote{also Center of Theoretical Nuclear Physics,
 National  Laboratory of Heavy Ion Accelerator,
Lanzhou 730000, China}\\  Institute of Nuclear Research,\\
  Chinese Academy of Sciences, Shanghai 201800, China\\
 }

\date{}
\begin{document}
\maketitle
\bigskip
\baselineskip 20.6pt
\begin{abstract}
\baselineskip 18pt
 The thermodynamic interaction at thermodynamic
equilibrium in the free fermion gas is alternatively described by
the coupling of particles with a scalar thermodynamic field which
features a self-interaction.  The gauge fields in $SU_c(3)$
symmetry are investigated with thermodynamic interactions. Six
off-diagonal gluons acquire effective masses through the
decolorization of thermodynamic fields using the Higgs mechanism.
Together with the rise of the fermion mass in the thermal bath,
this gives the partition of the thermodynamic energy under the
classical limit.

\thanks{PACS: 11.15.-q, 11.10.Ef,  51.30.+i }

\thanks{Keywords: Thermodynamic interactions, gauge fields, Lorentz scalar }

\end{abstract}

In thermodynamics, it was once a hot controversy  how  the
temperature transformed under the Lorentz boost (for a review, see
Ref.\cite{haar} and references therein).  Recently, the discussion
on the controversy  has been still going on (e.g., see
Refs.\cite{land,aldro,fenech}). This controversy may be
necessarily related to that how to define the proper temperature
and recognize the thermodynamic interaction in a covariant
framework for specific thermal systems. For fermion gases in a
covariant framework, it is economic to begin with the problem in
the relativistic field theory based on the following
consideration. The non-linearity as well as non-inertia due to
bremsstrahlung processes and multiple collisions in the thermal
system is beyond the kinetics given by equations of motion for
free fermions. Some dynamical degrees of freedom need to be
introduced. According to the notion of  field theories, the
interaction is mediated by intermediate bosons. One may therefore
introduce appropriate intermediate bosons (understood as the
thermodynamic fields) to describe thermodynamic interactions for
the fermion gas at thermodynamic equilibrium in the relativistic
framework. To do so, it is necessary to begin with two senarios
about the proper mass at thermal environment and thermodynamic
equilibrium. The first senario (I) is that the mass modified by
the heat  carried by the system is still observed as the proper
mass at the rest frame, which abides by the spirit to define the
proper mass. The second senario (II) is that the thermodynamic
equilibrium is Lorentz invariant, which was once adopted in Ref.
\cite{aldro}.

The perfect fermion gas which is free of interactions is a simple
system. Beyond that, it is convenient to investigate the
interactions in the fermion gas by introducing the covariant
derivative instead of the conventional derivative, as the
interactions of fermions are fixed by the gauge
invariance\cite{yang}.  The exact gauge invariance does not allow
the mass term of gauge fields. It is important to see whether the
spontaneous symmetry breaking occurs for the thermodynamic field
and then what will happen for gauge fields with thermodynamic
interactions which exist for fermions with various internal
degrees of freedom (charges), considering that  the gauge boson
acquires mass through the coupling with the field which displays
the spontaneous breakdown of symmetry under the internal Lie
group\cite{higgs,abers}. The exploration of possible mass
acquisition for gauge bosons with thermodynamic interactions is of
significance, for instance, in the possible color-deconfined
system of quark-gluon plasma (QGP).

Note that the system we study here is at thermodynamic
equilibrium, the equipartition of thermodynamic energy will become
the only soluble condition for the proposed idea. This restricts
ourselves to the topic under the classical limit, and thus it is
quite different from the past investigation of  the screening mass
in the propagator of quantized gauge fields in the thermal bath
(for monographs, see Ref.\cite{bellac}). The  thermodynamic energy
partition explored in this study may be applied to analyze the
initial temperature of QGP that is important for the QGP evolution
(e.g., see Refs.\cite{wang,qgp}).  In the following, thermodynamic
interactions in the free fermion gas are described in the
relativistic field theory at first. An investigation of $SU_c(3)$
gauge bosons in the interacting thermal fermion gas follows then.

The most favorable bosons are the scalar and vector ones. The
vector boson couples to the non-zero conserved thermodynamic
current which does not exist in the system at thermodynamic
equilibrium. One may argue that the temporal component of the
vector boson may exist. But such an existence of the temporal
component contradicts with the Senario (II), since the thermal
current in the thermal body appears as performing the Lorentz
boost. Consequently, only the scalar boson mediates the
thermodynamic interaction. The coupling of the scalar boson with
the fermion is equivalent to the heating which increases the
fermion mass. This requires the imaginary mass of the scalar
boson, based on the knowledge of the mean-field treatment. The
boson with the imaginary mass is however unphysical, and the
self-interaction has to be introduced so as to obtain the real
mass. The spirit in doing so is similar to that for the
pseudoscalar field\cite{gold} and for the Higgs scalar
field\cite{higgs}.

The scalar boson field,  noted as the thermodynamic field below,
is described by the following Lagrangian
\begin{equation}
{\cal L}_\tau=\frac{1}{2}
(\pp_\mu\phi_\tau\pp^\mu\phi_\tau-\mu_\tau^2\phi_\tau^2)
-\frac{\ld}{4}\phi_\tau^4\label{lag1}
\end{equation}
where  $\mu_\tau$ and $\ld$ are the constants with $\mu_\tau^2<0$
and $\ld>0$.  Yet, we need to obtain the real mass of $\phi_\tau$.
The vacuum solution of field $\phi_\tau$ is obtained as
${\phi_\tau^0}^2=-\mu^2_\tau/\ld $, by minimizing the potential.
The constant vacuum field is symmetry-breaking under the
reflection of internal field space. Substituting
\begin{equation}
\phi_\tau=\phi_\tau^0+\phi=\sqrt{-\mu^2_\tau/\ld}+\phi
\end{equation}
into Eq.\ref{lag1}, it becomes
\begin{equation}
{\cal L}_\tau=
\frac{1}{2}(\pp_\mu\phi\pp^\mu\phi-m_\tau^2\phi^2)-\ld
\phi_\tau^0\phi^3 -\frac{\ld}{4}\phi^4 \label{lag2}
\end{equation}
where the real mass of $\phi$ is obtained to be
$m_\tau=\sqrt{-2\mu^2_\tau}$. Now, the vacuum of the thermodynamic
field is actually redefined and  is  symmetric under field
reflection, while the reflection symmetry of ${\cal L}_\tau$ is
broken due to the term $-\ld \phi_\tau^0\phi^3$. The scalar
thermodynamic field defined here is in form similar to the Higgs
field, and its spontaneous-symmetry-breaking property under
internal Lie groups can be found in Refs.\cite{higgs,abers,gold}.

The total Lagrangian which contains the fermion,  thermodynamic
field, and the Yukawa coupling between them is
 \begin{equation}
  {\cal L} =\bar{\psi}(i\gm_\mu \pp^\mu -m-g_\tau \phi_\tau)
\psi+ {\cal L}_\tau
 \label{lag3}
 \end{equation}
where $\psi$ is the Dirac spinor, and  $m$ is the proper mass of
fermion. In homogeneous thermal matter, the scalar field $\phi$ is
given as
\begin{equation}
\phi=\frac{1}{m^2_\tau}(-g_\tau\bar{\psi}\psi-3\ld\phi_\tau^0\phi^2
-\ld\phi^3) \label{field2}
\end{equation}
In an alternative description for the thermodynamic interaction,
the real-mass boson does not always suggest the definite existence
of the boson. Actually, a scalar boson which serves the role like
Higgs bosons is not found yet. Alternatively, one may take the
boson mass $m_\tau$ as large enough. In this way, the field $\phi$
will be as small enough. This treatment is consistent with the
equipartition theorem, as seen below. Now, the thermal heat of the
system may just be determined by the term
$g_\tau\bar{\psi}\phi_\tau^0\psi$ in Eq.\ref{lag3}. The
homogeneous thermodynamic equilibrium of the fermion system can be
elaborated by the coupling with the isotropic field vacuum
$\phi_\tau^0$. Thus, it is actually the broken vacuum of the
scalar field that plays the very role, in analogy to that of the
Higgs field in gauge theories\cite{higgs,abers}.

The potential $g_\tau\phi_\tau^0$ is a homogeneous quantum for all
particles in the mean-field approximation, which indicates  the
equipartition theorem that holds for the classical limit is the
unique solvable condition for the problem. The fermion mass
measured at the rest frame at the non-relativistic limit is thus
\begin{equation}
m^*=m+g_\tau\phi_\tau^0=m+\frac{3}{2}T_0 \label{eqt3}
\end{equation}
The derivation of Eq.\ref{eqt3} implies that it is reasonable to
take $m_\tau$ as large enough. Otherwise,  it will lead to the
inequality $m^*\neq m+3/2T_0$ for finite $m_\tau$ not large
enough. The proper temperature $T_0$ in the free fermion gas,
which is proportional to the modification of the proper mass of
fermions as given in Eq.\ref{eqt3}, is a Lorentz scalar,
consistent with the Senario (I).  It may be straightforward to
work out various Lorentz transformations for the apparent
temperature based on the Lorentz structure of Lagrangian.

The thermodynamic interaction has been alternatively described
through the dynamic scalar boson coupling in the free fermion gas
in above. Now, let's investigate the more realistic system, the
interacting fermion gas. The interaction forms of the interacting
fermions are fixed by the gauge invariance\cite{yang}. The
applications of gauge theories can be found, for instance, in
Refs.\cite{abers,mori}. In the following, we consider the
thermodynamic interaction in the color $SU_c(3)$ symmetry. The
interactions in the Lagrangian for thermal fermions are introduced
by replacing the partial derivative $\pp_\mu$ with the covariant
derivative $D_\mu=\pp_\mu-igA_\mu$ with $g$ the coupling constant.
The gauge field is expressed with respect to the generators $\ld$
in matrix form as $A_\mu=\ld^\af A_\mu^\af/2$, with
$\af=1,\cdots,8$ in $SU_c(3)$ symmetry.  The classical Lagrangian
involving the gauge field can be given as
 \bey {\cal L} &=&\bar{\psi}(i\gm_\mu D^\mu -m-g_\tau \phi_\tau)
\psi\nnb\\ &&+Tr\{\frac{1}{2}[(D_\mu\phi_\tau)^\dagger
D^\mu\phi_\tau-\mu_\tau^2(\phi_\tau^\dagger\phi_\tau)\} \nnb\\
 &&-\frac{\ld}{4}(Tr[\phi_\tau^\dagger\phi_\tau])^2
 -\frac{1}{4}F^{\af}_{\mu\nu} F^{\af\mu\nu} \label{lag5}\eey
where  the trace is over the internal color space.
Mathematically, there can be the terms linear in $Tr\phi_\tau$,
$Tr\phi_\tau^3$, and $Tr(\phi_\tau^\dagger\phi_\tau)^2$. Assuming
the reflection symmetry for the thermodynamic field $\phi_\tau\ra
-\phi_\tau$, the terms having the odd powers are ruled out.
Without involving the term linear in
$Tr(\phi_\tau^\dagger\phi_\tau)^2$, one may obtain the field
vacuum
\begin{equation}
\phi_\tau^0=<\sqrt{Tr(\phi_\tau^\dagger\phi_\tau)}>_0=
\sqrt{-\mu_\tau^2/\ld} \label{vac}
\end{equation}
which is a natural extension of the vacuum of $O(1)$ symmetry
given by Eq.\ref{lag3}. In this way, it is simple to perform the
rotation in the internal space without adding any additional
assumption on the vacuum value of thermodynamic fields. In order
to keep the hermitian constraint for the Lagrangian,  the field
$\phi_\tau$ should be hermitian, i.e.
$\phi_\tau^\dagger=\phi_\tau$. The gauge field strength tensor is
\begin{equation}
F^{\af}_{\mu\nu}=\pp_\mu A_\nu^\af-\pp_\nu A_\mu^\af+ g
f^{\af\bt\gm}A_\mu^\bt A_\nu^\gm
\end{equation}
with $f^{\af\bt\gm}$ the antisymmetric structure constants. The
covariant derivatives for the fermion and thermodynamic field are
defined, respectively
\begin{eqnarray}
D_\mu\psi&=&(\pp_\mu-igA_\mu^\af\ld^\af/2)\psi\nnb\\
D_\mu\phi_\tau&=&\pp_\mu\phi_\tau-i\frac{g}{2}
 [A_\mu^\af\ld^\af,\phi_\tau]
 \label{gaug}
\end{eqnarray}
Under the gauge transformation
 \bey
&&\psi^\prime=S^{-1}\psi,\hbox{ }\phi_\tau^\prime=S^{-1}\phi_\tau
S\nnb\\ && A_\mu^\prime=S^{-1}A_\mu S+\frac{i}{g}S^{-1}\pp_\mu S
\label{gauge}
 \eey
where $S=\exp[i\theta^\af(x)\ld^\af/2]$ with $\theta^\af(x)$ the
gauge functions, the Lagrangian (\ref{lag5}) is invariant.

Because thermodynamic interactions exist for fermions with various
charges (colors), the hermitian matrix of thermodynamic fields is
not required to be diagonalized. Similar to the non-interacting
case, the apparent consequence resulted from the thermodynamic
interaction is the modification of the fermion mass. We therefore
need to perform the diagonalization for the $\phi_\tau$. The
hermitian matrix can be diagonalized through the unitary
transformation as follows
\begin{equation}
\phi_\tau=\exp[i\frac{\xi^\af\ld^\af}{2}]
\left (\begin{array}{ccc}
 \phi_{\tau1} & & \\ &\phi_{\tau2}& \\ & & \phi_{\tau3}
\end{array}\right)\exp[-i\frac{\xi^\af\ld^\af}{2}]
\label{diag}
\end{equation}
In diagonalized form, the $\phi_\tau$ vacuum quantity is
$\phi_\tau^0=\sqrt{\sum_i\phi_{\tau i}^{02}}$, and one may choose
any of following forms to express the vacuum in the tensor form
\begin{equation}
v=\left (\begin{array}{ccc}
 \phi_\tau^0 & & \\ &0& \\ & & 0
\end{array}\right),
 \left (\begin{array}{ccc}
  0& & \\ &\phi_\tau^0& \\ & & 0
\end{array}\right),
\left (\begin{array}{ccc}
  0& & \\ &0& \\ & & \phi_\tau^0
\end{array}\right)\label{dirac5}
 \end{equation}
Any one of tensor vacua can be rotated to other forms by the
symmetry-breaking generator $\ld^\bt$ such as performing $\ld^\bt
v_{\bt\bt} \ld^\bt$. The thermodynamic field can be parametrized
based on the given vacuum. The parametrization is simple as
knowing that the diagonal elements in Eq.\ref{diag} are the real
quantities. We just need replace the medium matrix in
Eq.\ref{diag}, and it is
\begin{equation}
\phi_\tau=\exp[i\frac{\xi^\af\ld^\af}{2}](u+v)
\exp[-i\frac{\xi^\af\ld^\af}{2}] \label{diag1}
\end{equation}
where $u$ is the same as $v$ but the nonzero vacuum quantity is
substituted by the field $\phi$. In order to give the modification
to the fermion mass of all colors by thermodynamic interactions,
three forms of vacuum are all needed. Thus the Lagrangian should
involve thermodynamic fields with all of three vacuum choices, and
for simplicity  it is not rewritten here.

To give the modification to the fermion mass, the exponential
factors in Eqs.\ref{diag},\ref{diag1} should be eaten. Moreover,
massless Goldstone bosons\cite{gold}, appearing as the symmetry of
vacuum is broken under the present $SU_c(3)$ group, should be
exorcised using the Higgs mechanism\cite{higgs}. In the
parametrized expression of $\phi_\tau$ given in Eq.\ref{diag1},
Goldstone bosons are described by tensors $[\ld^\bt,v]\xi^\bt$
which satisfy the relations $\ld^\bt v\neq 0$ and
$v\ld^\bt=(\ld^\bt v)^\dagger \neq 0$ with $M$ $(<8)$ choices for
$\bt$. Both tasks are accomplished simultaneously as we perform
the gauge transformation given in Eq.\ref{gauge} under the unitary
gauge $\theta^\af=\xi^\af$.

A straightforward consequence of the coupling between the gauge
and thermodynamic fields is the generating of gauge boson masses,
abiding by the Higgs mechanism. The mass term of the gauge field
is obtained from the term $(D_\mu\phi)^\dagger D^\mu\phi$ in
Eq.\ref{lag5}. To acquire the effective gluon mass, we just need
sum up the terms linear in $v^2$ contained  in
$(D_\mu\phi)^\dagger D^\mu\phi$ over three vacua as
\begin{equation}
m^{2}_\af A_\mu^\af A^{\af\mu}=-\frac{g^2}{4}\sum_v Tr\{[\ld^\af
A_\mu^\af, v] [\ld^\sg A_\mu^\sg, v]\}
\end{equation}
with $\af,\sg=1,\cdots,8$. Thus, the effective gluon masses are
\begin{equation}\label{mg}
m_\af=g\phi_\tau^0, \hbox{ for } \af\neq 3,8
\end{equation}
For $A_\mu^3$ and $A_\mu^8$, they keep massless.

An important procedure adopted above is the diagonalization which
is actually the decolorization of thermodynamic fields.   Through
the decolorization of thermodynamic fields, six off-diagonal
gluons acquire effective masses.  For the electromagnetic
interaction in U(1) symmetry, the coupling of fermions with the
thermal bath is through the exchange of the neutral thermodynamic
field, according to the hermitian constraint on the Lagrangian.
There is no coupling between the gauge boson and thermodynamic
field in $U(1)$ symmetry where there is the relation
$D_\mu\phi_\tau=\pp_\mu\phi_\tau$ for the thermodynamic field as
given by Eq.\ref{gaug}, and this is actually quite general  for
the symmetry in abelian groups. No photo can acquire the effective
mass in the thermodynamic field.  This is consistent with the fact
that the equipartition theorem is not able to be applied to the
photos for which the quantum effect is always important. Similar
to photos, two diagonal colorless gluons which relate to the color
neutral current keep massless. The mass acquisition of
off-diagonal gauge bosons in other $SU(N)$ symmetries with
thermodynamic interactions may be on the analogy of the present
case in $SU_c(3)$ symmetry. As known from above, the effective
gluon mass is linear in the proper temperature if the temperature
is defined by $g_\tau\phi_\tau^0$ as in Eq.\ref{eqt3}.

The equipartition theorem is appropriate for the limit that
particles may pass through all possible states. As the particles
are in high degeneracy or the quantum effect is important,
particles are not able to pass through all states. Thus, the
quantity $g_\tau\phi^0_\tau$ provided in the mean-field
approximation exists on the sense that the equipartition theorem
is appropriate for the thermal system. For the gauge bosons, the
acquisition of the effective mass is equivalent to the partition
of the thermodynamic energy at the classical limit.  So, the
classical Boltzmann distribution can be applied to the quarks and
six off-diagonal gluons for some cases that are consistent with
the senario (I), whereas for the photon and colorless gluons the
quantum effect is always important. If the equipartition theorem
is applicable, the effective masses of colored gluons is obtained
as $m_\af=3T_0$ for the ultra-relativistic case, and  same
effective quark masses are obtained considering the interaction is
weak due to asymptotic freedom. This case may eventually exist in
an extremely hot QGP with a high fugacity where the quantum effect
in distributions of colored partons may be negligible\cite{wang}.
However this is not simply to say that it has the relation
$g_\tau=g$ since here the interactions in QGP is neglected. In
addition to considering photons and diagonal gluons always obey
the Bose distribution, one is able to carry out the partitioned
energy for each parton in the sense of statistical average. As the
quantum effect in parton distributions becomes important during
the cooling, the concept of the equipartition is not applicable
since the universal mean quantity ($g_\tau\phi_\tau^0$) does not
exist again.

In summary, the thermodynamic interaction in fermion gases is
alternatively described by the exchange of a scalar boson,  based
on notions of the relativistic field theory and the senario for
thermodynamic equilibrium.  The scalar boson field (thermodynamic
field) features a self-interaction like the Higgs scalar field.
The gauge fields in $SU_c(3)$ symmetry are investigated with
thermodynamic interactions. The coupling between the gauge and
thermodynamic fields results in the generation of the effective
mass of six off-diagonal gluons, which is realized through the
decolorization of thermodynamic fields using the Higgs mechanism.
Two diagonal gluons still keep massless. Apart from the nonabelian
case, the mass acquisition does not take place for the abelian
gauge boson in thermodynamic interactions. The modification of the
fermion proper mass in the thermal bath is equivalently described
by the coupling with the symmetry-breaking vacuum of the scalar
thermodynamic field.  The effective  masses acquired are the
partitioned thermodynamic energies that are observed to have the
same property of the proper mass in the center of mass frame. The
investigation holds for the case that the quantum effect together
with the interaction of the system is weak.

\section*{Acknowledgement}
This work is partially supported by CAS Knowledge Innovation
Project No.KJCX2-N11  and the Major State Basic Research
Development Program under grant No. G200077400.

\end{document}